\documentclass[epj]{webofc}
\usepackage[utf8]{inputenc}
\usepackage[varg]{txfonts}   
\usepackage{booktabs}
\usepackage{xcolor}
\definecolor{darkred}{rgb}{0.4,0.0,0.0}
\definecolor{darkgreen}{rgb}{0.0,0.4,0.0}
\definecolor{darkblue}{rgb}{0.0,0.0,0.4}
\usepackage[bookmarks,linktocpage,colorlinks,
    linkcolor = darkred,
    urlcolor  = darkblue,
    citecolor = darkgreen]{hyperref}
\wocname{EPJ Web of Conferences}
\woctitle{Lattice2017}
\newcommand{\norm}[1]{\left\lVert#1\right\rVert}
\newcommand{\abs}[1]{\left|#1\right|}
\newcommand{\pdagger}{{\phantom{\dagger}}}
\newcommand{\ND}{\text{\scriptsize{ND}}}
\begin{document}
%
\selectlanguage{english}
\title{%
	Multigrid accelerated simulations for Twisted Mass fermions
}
\author{%
\firstname{Simone} \lastname{Bacchio}\inst{1,2}\fnsep\thanks{Speaker, \email{s.bacchio@hpc-leap.eu}}
\and \firstname{Constantia} \lastname{Alexandrou}\inst{1,3}
\and \firstname{Jacob}  \lastname{Finkerath}\inst{3} 
}
\institute{%
Department of Physics, University of Cyprus, PO Box 20537, 1678 Nicosia, Cyprus
\and
Fakult\"at f\"ur Mathematik und Naturwissenschaften, Bergische Universit\"at Wuppertal,
Wuppertal, Germany
\and
Computation-based Science and Technology Research Center, The Cyprus Institute, Nicosia, Cyprus
}
\abstract{
Simulations at physical quark masses are affected by the critical slowing down of the solvers.
Multigrid preconditioning has proved to deal effectively with this problem. 
Multigrid accelerated simulations at the physical value of the pion mass are being performed  to generate $N_f = 2$ and $N_f = 2 + 1 + 1$ gauge ensembles using  twisted mass fermions. 
The adaptive aggregation-based domain decomposition multigrid solver, referred to as DD-$\alpha$AMG method, is employed for these simulations. 
Our simulation strategy  consists of an hybrid approach of different solvers, 
involving the Conjugate Gradient (CG), multi-mass-shift CG and DD-$\alpha$AMG solvers. 
We present an analysis of  the multigrid performance during the simulations discussing the stability of the method.
This significant speeds up the Hybrid Monte Carlo simulation by more than a factor $4$ at physical pion mass compared
to the usage of the CG solver.
}
\maketitle

\section{Introduction}\label{intro}
Simulations at the physical value of the pion mass have been intensively pursued by a number of lattice QCD collaborations. In order to accomplish these simulations speeding up of the solvers is an essential step.
A successful approach employed is 
based on multigrid methods used in preconditioning standard Krylov solvers.
There is a number of variant formulations of highly optimized multigrid solvers, which yield improvements of more 
than an order of magnitude for operators at the physical value of the light quark mass, 
as reported in Refs.~\cite{Babich:2010qb, Frommer:2013fsa, Alexandrou:2016izb, Clark:2016rdz}.

In this work, we focus on simulations with twisted mass (TM) fermions.  
This discretization scheme has the advantage that all observables
are automatically ${\cal O} (a)$ improved when tuned at maximal twist~\cite{Frezzotti:2003ni}.
This formulation is thus particularly suitable for hadron structure studies, since the probe, such as the axial current, needs no further improvement in contrast 
to clover improved Wilson Dirac fermions. Furthermore, the presence of a finite TM term bounds the spectrum of 
$DD^\dagger$ from below by a positive quantity $\mu^2$, where $D$ is the Wilson Dirac operator and $\mu$ is the TM parameter. 
This avoids exceptional configurations and, at the same time, gives an upper bound to the condition number, satisfying 
the convergence of numerical methods. Using this discretization approach enables
us to study a wide range of
observables. Both these simulations and the  calculation of observables
have substantially benefit from employing multigrid methods.

Here we will show results for two simulations at maximal twist and at the physical value of the pion mass, which have been generated in the last two years using twisted mass fermions.
The properties of these ensembles are listed in Table~\ref{tab:ensembles}.
Results of the $N_f=2$ simulation have been partially presented in Refs.~\cite{Alexandrou:2016izb,Bacchio:2016bwn}, where a statistic of 2000 molecular dynamic units (MDU) has been used.
The tuning procedure of the $N_f=2+1+1$ ensemble and some physical results are
presented in  Ref.~\cite{Finkenrath:2017}. 
\begin{table}[t]
	\centering
	\small
	\begin{tabular}{l | c c c c}
		\toprule
		 & $V$ & $a$~[fm] & $m_\pi$~[MeV] & MDUs \\
		\midrule
		$N_f=2$ & $64^3\times128$ & $0.0936(5)$ & $135.2(7)$ & 3128\\
		$N_f=2+1+1$ & $64^3\times128$ & $0.0820(4)$ & $136.1(7)$ & 3051\\
		\bottomrule
	\end{tabular}
	\caption{\label{tab:ensembles} We give the lattice volume $V$, the lattice spacing $a$, the pion mass $m_\pi$
	and the number of molecular dynamics units (MD) measured in MD trajectories of the two ensembles discussed in this work. }
\end{table}

Here we discuss in detail  the HMC simulations 
with a focus on the usage of multigrid methods and in particular
the performance of the
DD-$\alpha$AMG solver adapted for TM fermions~\cite{Alexandrou:2016izb}. 
We discuss our strategy for the calculation of the force terms
and the generation of the multigrid subspace during the integration of the molecular
dynamics (MD) and demonstrate that this  yields stable simulations with 
an improved performance.

\subsection{DD-$\alpha$AMG method}
The adaptive aggregation-based domain decomposition multigrid (DD-$\alpha$AMG) method 
has been introduced in Ref.~\cite{Frommer:2013fsa} as a solver for the clover-improved Wilson Dirac operator $D$. 
In the DD-$\alpha$AMG method a flexible iterative Krylov solver is preconditioned at every iteration step by a multigrid approach given by the error propagation
\begin{equation}\label{eq:error_propagation}
\epsilon\,\leftarrow\,\left(I-MD\right)^k\left(I-PD_{c}^{-1}P^\dagger D\right)\epsilon,
\end{equation}
where $M$ is the smoother, $k$ are the number of smoothing iterations, $P$ is the prolongation operator and $D_{c} = P^\dagger D P$ is the coarse Wilson operator.
The multigrid preconditioner exploits domain decomposition 
strategies having for instance as a smoother the Schwartz Alternating Procedure (SAP)~\cite{Luscher:2003qa}
and as a coarse grid correction an aggregation-based coarse grid operator. 
The method is designed to deal efficiently with both, infrared (IR) and ultra-violet (UV) modes of $D$. 
Indeed, the smoother reduces the error components belonging to the UV-modes~\cite{Frommer:2013fsa}, while the
coarse grid correction deals with the IR-modes. This is achieved by using an interpolation operator $P$, which approximately 
spans the eigenspace of the small eigenvalues. 
Thanks to the property of local coherence~\cite{Luscher:2007se} 
the subspace can be approximated by aggregating over a small set of $N_v\simeq\mathcal{O}(20)$ test vectors $v_i$,
which are computed via an adaptive setup phase~\cite{Frommer:2013fsa}.
We remark that the prolongation operator in the DD-$\alpha$AMG method
is $\Gamma_5$-compatible, i.e.~$\Gamma_5P = P\Gamma_{5,c}$.
Thanks to this property the $\Gamma_5$-hermiticity of $D$ is
preserved in the coarse grid as well, i.e.~$D_c^\dagger = \Gamma_{5,c}D_c\Gamma_{5,c}$.

\begin{figure}[htb]
	\centering
	\includegraphics[width=0.75\linewidth]{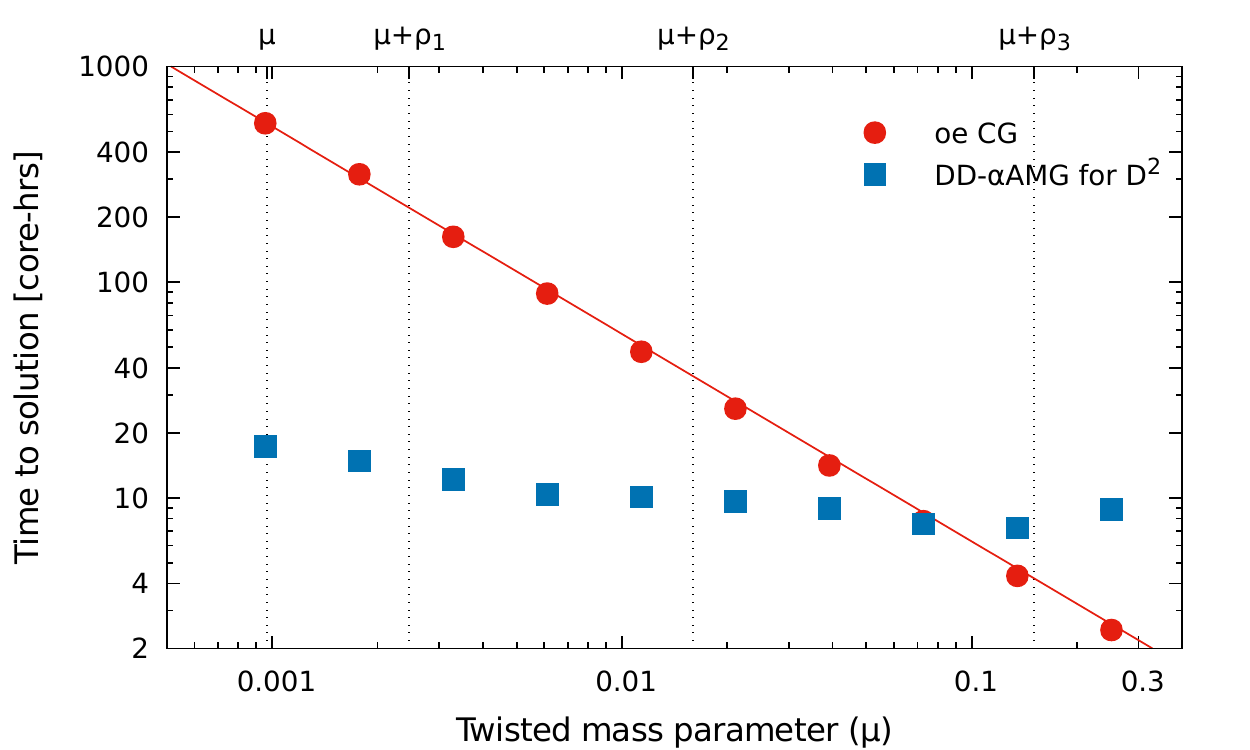}
	\caption{\label{fig:crit_slow_down} Comparison of the time to solution for inverting the squared operator 
	$D^\dagger(\mu)D(\mu)$ using either an odd-even (oe) CG solver or a three level DD-$\alpha$AMG solver. 
	In the DD-$\alpha$AMG method the inversion is done in two steps inverting first $D(-\mu)$ and then $D(\mu)$.
	The results are for the $N_f=2$ ensemble listed in Table~\ref{tab:ensembles}. The shifts $\rho_i$ are the values used for the Hasenbusch mass preconditioning in the HMC simulation.}
\end{figure}

The DD-$\alpha$AMG approach has been adapted in Ref.~\cite{Alexandrou:2016izb} to the Wilson TM operator $D(\pm\mu) = D \pm i \mu \Gamma_5$.
Due to the $\Gamma_5$-compatibility, the coarse operator reads similarly to the fine operator, i.e. $D_c(\pm\mu) = D_c \pm i \mu \Gamma_{5,c}$, and the same prolongation operator $P$ can be used for both signs of $\mu$.
In $N_f=2+1+1$ simulations we use a three level DD-$\alpha$AMG solver also for the non-degenerate TM operator 
\begin{equation}\label{eq:non-degenerate}
D_{\ND}(\bar\mu,\bar\epsilon)= (D\otimes I_2)+i\bar{\mu}\,(\Gamma_5\otimes\tau_3)-\bar{\epsilon}\,(I\otimes\tau_1)=
\begin{bmatrix}
D(\bar\mu)        & -\bar\epsilon\,I \\
-\bar\epsilon\,I & D(-\bar\mu)
\end{bmatrix} 
\end{equation}
where $I_2$, $\tau_1$ and $\tau_3$ act in flavor space. Here the coarse operator is constructed  with the same prolongation operator $P$ of $D(\pm\mu)$ although it is used diagonally in flavor space, i.e. $P_\ND = P\otimes I_2$. 
Thus, the coarse non-degenerate TM operator is defined as $D_{\ND,c}(\bar\mu,\bar\epsilon) = P_\ND^\dagger D_{\ND}^\pdagger(\bar\mu,\bar\epsilon) P_\ND^\pdagger$.

In Fig.~\ref{fig:crit_slow_down} we report the comparison of time to solution between CG and DD-$\alpha$AMG solvers when the squared operator 
$D^\dagger(\mu) D(\mu)$ is inverted at different values of $\mu$.
These inversions are computed with the DD-$\alpha$AMG solver in two steps solving as first $D(-\mu)\,x =\Gamma_5 b$ and then $D(\mu)\,y =\Gamma_5 x$.
Thus, the computational cost for solving $D^\dagger(\mu) D(\mu)\,y = b$ is double as compared to the cost for inverting $D(\pm\mu)$.
At the physical value of the light quark mass the DD-$\alpha$AMG solver gives a speed-up of more than two order of magnitude compared to CG solver.
When it is used for the non-degenerate TM operator at the physical strange quark mass
the speed-up is around one order of magnitude.

\subsection{Characteristics of the  simulations}

The simulations  hare produced by using the tmLQCD software 
package~\cite{Jansen:2009xp} and the DDalphaAMG library for TM fermions~\cite{Bacchio:2016bwn}.
All the simulation codes are released under GNU license.
The integrator is given by a nested minimal norm scheme of order 2~\cite{Omelyan:2003272} with a nested 
integration scheme setup similar to previously produced simulations~\cite{Abdel-Rehim:2015pwa}.
We used a three-level DD-$\alpha$AMG method for the small mass terms, a CG solver for larger
mass terms and a combination of a multi-mass shift CG solver and  three-level DD-$\alpha$AMG method to compute
the rational approximation. The simulations are performed for the even-odd reduced operator.
For  the heat-bath inversions and acceptance steps we require  as stopping criterion for the solvers
the relative residual to be smaller than $10^{-11}$. 
For the force terms in the MD trajectory the criterion is relaxed,
using $10^{-7}$ for the CG solver and $10^{-9}$ for the DD-$\alpha$AMG method.
This ensures that the reversibility violation of the MD integration is
sufficiently reduced. Note that the usage of the multigrid method is efficient
if the subspace can be reused  at larger integration time.
In general, this yields a larger reversibility violation.
By choosing a higher accuracy i.e. by using a smaller stopping criterion,
for the multigrid solver the reversibility violation can be reduced. We check that with the values mentioned above the reversibility
violations are compatible with the case where a CG solver is used for all the monomials.

\section{$N_f=2$ multigrid accelerated HMC simulations}
\begin{figure}
	\centering
	\includegraphics[width=0.9\linewidth]{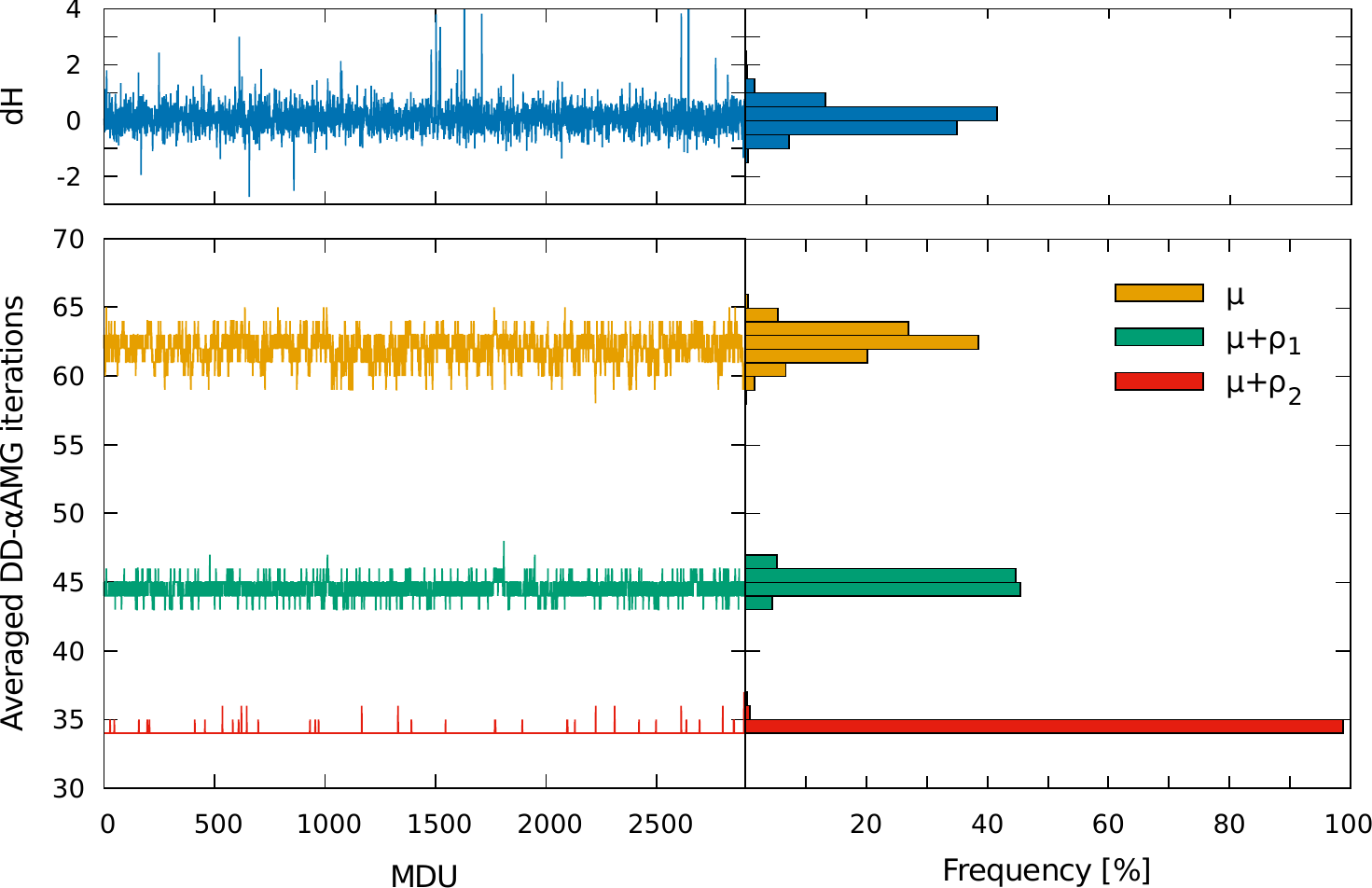}
	\caption{\label{fig:iter_nf2} {Top:} The energy violation of the numerical integrator used during the MD of the $N_f=2$ ensemble
	 plotted in units of MD trajectories (left) and its frequency (right).
	{Bottom:} DD-$\alpha$AMG iterations count averaged on MDU considering different monomials of 
	the HMC simulation  shown in units of MD trajectories (left) and its frequency (right). The results are for the $N_f=2$ ensemble listed in Table~\ref{tab:ensembles}. The shifts $\rho_i$ 
	are \{0.0015, 0.015\} and $\mu=0.000966$. The iterations count corresponds to average of the sum of the count of the outer level during the calculation of the force in the MD.}
\end{figure}

In the $N_f=2$ simulations we employ the Hasenbusch mass preconditioning~\cite{Hasenbusch:2002ai} by introducing 
additional mass terms and split up the determinants into additional ratios as in the following
\begin{equation}\label{eq:Hasenbusch}
\det\left[Q^2+\mu^2\right]=\det\left[\frac{Q^2+\mu^2}{Q^2+(\mu+\rho_1)^2}\right] \det\left[\frac{Q^2+(\mu+\rho_1)^2}{Q^2+(\mu+\rho_2)^2}\right]\det\left[\frac{Q^2+(\mu+\rho_2)^2}{Q^2+(\mu+\rho_3)^2}\right]\det\left[Q^2+(\mu+\rho_3)^2\right]
\end{equation}
where $Q=\Gamma_5 D = Q^\dagger$ is the hermitian Wilson Dirac operator. 
Each determinant in the right hand side (rhs) of Eq.~\ref{eq:Hasenbusch} can be placed on a 
different monomial and integrated on a different time-scale.
The time-steps are chosen accordingly to the intensity of the force term.
This procedure  controls  the large fluctuations of the force terms, avoiding instabilities during the HMC.

In our $N_f=2$ simulation given in Table~\ref{tab:ensembles}, we  use 5 time-scales, which are integrated respectively $N_{\text{int}}=\{20, 60, 180, 540, 1620\}$ times.
The gauge action is placed in the innermost time-scale; in the other time-scales we place one by one the fermionic determinants from the rhs of Eq.~(\ref{eq:Hasenbusch}) going from the largest shift to the smallest.
As depicted in Fig.~\ref{fig:iter_nf2}, this yields a stable simulation without large spikes in the energy violation $\delta H$ with an acceptance rate of 84.5\%.
For the Hasenbusch mass preconditioning we use the shifts $\mu+\rho_i$ depicted in Fig.~\ref{fig:crit_slow_down}. 
The DD-$\alpha$AMG method is faster than CG solver for all the shifts except the largest given by $\rho_3$.
Thus we have used DD-$\alpha$AMG for the inversions with shifts $\mu$, $\mu+\rho_1$ and $\mu+\rho_2$.
The DD-$\alpha$AMG iterations count averaged per MD trajectory is depicted in Fig.~\ref{fig:iter_nf2}.
No exceptional fluctuations or correlations with larger energy violation $\delta H$ are seen along the simulation.
The stability of the multigrid method is ensured by updating the setup every time the inversion at the physical quark mass, i.e.~$(Q^2+\mu^2)^{-1}$, is performed.
The update is based on the previous setup by using one setup iteration, which is possible due to the adaptivity in the DD-$\alpha$AMG method~\cite{Frommer:2013kla} for the definition of the setup iteration.
At the beginning of the trajectory we perform three setup iterations.
The final speed-up, including setup costs, is a factor of 
8 compared to CG in $N_f=2$ simulations at the physical pion mass.

\section{$N_f=2+1+1$ multigrid accelerated (R)HMC simulations}
\begin{figure}
	\centering
	\includegraphics[width=0.9\linewidth]{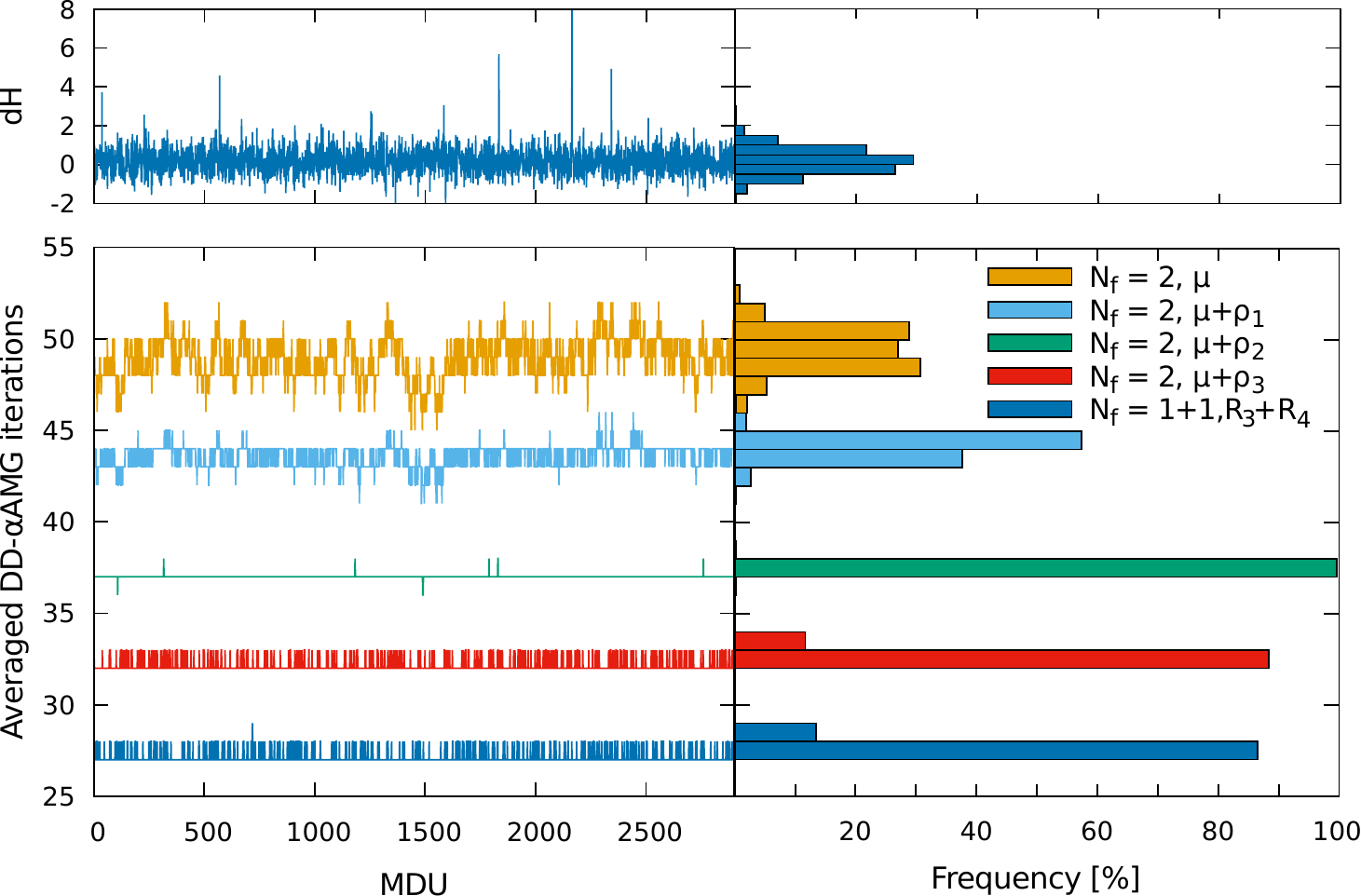}
	\caption{\label{fig:iternf211} {Top:} The energy violation of the numerical integrator used during the MD of the $N_f=2+1+1$ ensemble
		plotted in units of MD trajectories (left) and its frequency (right).
		{Bottom:} DD-$\alpha$AMG iterations count averaged on MDU considering different monomials of 
		the HMC simulation  shown in units of MD trajectories (left) and its frequency (right). The results are for the $N_f=2+1+1$ ensemble listed in Table~\ref{tab:ensembles}. The shifts $\rho_i$ 
		are $\{0.0003,0.0012,0.01\}$ and $\mu=0.00072$.
		The monomials $\text{R}_3$ and $\text{R}_4$ contain the four smallest shifts of the rational approximation.
		The iterations count corresponds to average of the sum of the counts of the outer level during the calculation of the force in the MD.}
\end{figure}
In $N_f=2+1+1$ simulations we follow the same prescription reported in the previous section. 
Additionally to the determinants in the rhs of Eq.~\ref{eq:Hasenbusch}, 
the determinant of the non-degenerate $N_f=1+1$ TM operator $D_\ND$ in Eq.~(\ref{eq:non-degenerate}) is included
in the action. Monte Carlo algorithms require a positive weight, which can be retrieved for the non-hermitian operator $D_\ND$ by using the Rational HMC (RHMC)~\cite{Clark:2006fx}.
Here the determinant is rewritten as
\begin{equation}\label{eq:detRHMC}
\det\left[D_{\ND}^{\phantom{-1}}(\bar\mu,\bar\epsilon)\right] = \det\left[\sqrt{Q_{\ND}^2}\right] = \det\left[R^{-1}_\ND\right]\det\left[\abs{Q_\ND^{\phantom{-1}}}R_\ND^{\phantom{-1}}\right],
\end{equation}
where $Q_{\ND}^\pdagger = (\Gamma_5\otimes\tau_1) D_\ND^\pdagger = Q_{\ND}^\dagger$ is the hermitian version of $D_\ND$.
The term $R_{\ND}$ is the optimal rational approximation of
\begin{equation}\label{eq:rationalQnd}
\frac{1}{\sqrt{Q_{\ND}^2}}\simeq R_{\ND}\equiv R_{n,\epsilon}\left(Q_\ND^2\right) = a_{n,\epsilon} \prod_{i=1}^{n}\frac{Q_\ND^2+c_{n,\epsilon,(2i-1)}I}{Q_\ND^2+c_{n,\epsilon,2i}I},
\end{equation}
where $n$ is the order of the approximation and $\epsilon = \lambda_{\min}/\lambda_{\max}$ is fixed by the smallest and largest eigenvalue of $Q_{\ND}^2$, $\lambda_{\min}$ and $\lambda_{\max}$, respectively.
The coefficients $a_{n,\epsilon}$ and $c_{n,\epsilon,i}$ are given by the Zolotarev solution for the optimal rational approximation~\cite{zolotarev1877application}. 
The rational approximation is optimal because the largest deviation from the approximated function is minimal according to the de-Vall\'ee-Poussin’s theorem.
In general, one can relax the approximation of the square root by introducing a correction term $\det\left[\abs{Q_\ND^{\phantom{-1}}}R_\ND^{\phantom{-1}}\right]$
as in the rhs of Eq.~(\ref{eq:detRHMC}). 
It takes into account  the deviation from $\abs{Q_{\ND}}$ being  close to the identity as much as
the rational approximation is precise. For this reason we include it only in the acceptance step.

\begin{figure}
	\includegraphics[width=0.5\linewidth]{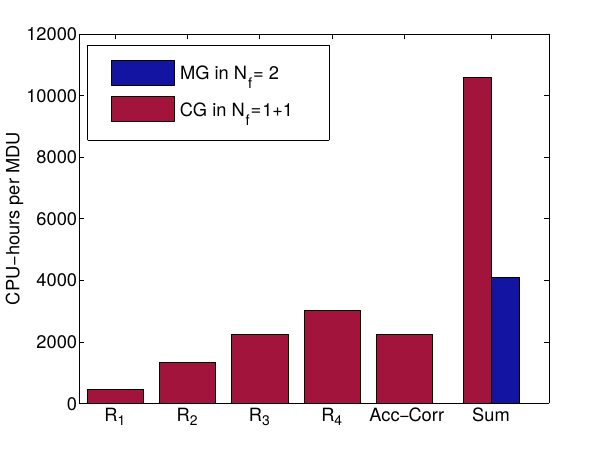}\hfill
	\includegraphics[width=0.5\linewidth]{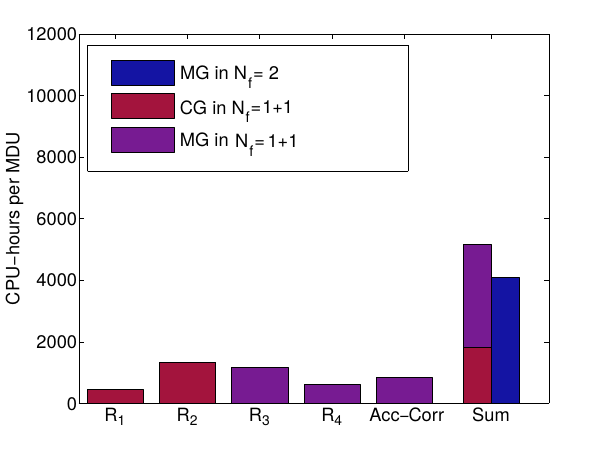}
	\caption{\label{fig:hmc_nf2+1+1} Costs per MDU of the $N_f=2+1+1$ simulation  shown in units of CPU hours.
		In the left panel all the inversions in the RHMC are computed with a MMS-CG solver while in the right panel
		an hybrid approach of MMS-CG and DD-$\alpha$AMG solvers is used.
		The last column, ``Sum'', reports the cost of RHMC in the $N_f=1+1$ sector compared to HMC in $N_f=2$ sector. The other columns regard different components 
		of the RHMC: the ``$\text{R}_i$''s are the rational approximation split in different monomial -- $\text{R}_1$ 
		contains the largest shifts while $\text{R}_4$ the smallest -- and ``Acc-Corr'' is the cost for the acceptance and correction step.}
\end{figure}

In our $N_f=2+1+1$ simulation with properties given in Table~\ref{tab:ensembles}, we use a rational approximation of order $n=10$, which has a relative deviation such that
$\norm{\abs{Q_{\ND}}R_\ND^{\phantom{-1}}}_\infty<1.4\cdot10^{-6}$,
considering the eigenvalues of $Q_\ND^2$ in the interval $\lambda_{\min}=6.5\cdot10^{-5}$ and $\lambda_{\max}=4.7$. 
The product of ratios in the rhs of Eq.~(\ref{eq:rationalQnd}) is split in four monomials $R_{1-4}$. 
The first two contain three shifts, the second two contain 
two shifts. We use a 6 level nested minimal norm second order integration scheme with 
integration steps $N_{\text{int}}=\{12, 36, 108, 324, 972, 2916\}$. The four monomials $R_{1-4}$ are placed one by one in the four outermost 
time-scales. 
As depicted in Fig.~\ref{fig:iternf211}, this yields relative small energy violation during the MD integration and an acceptance rate of 76.8\%.
In the same figure we  depict the iterations count for the inversions done with DD-$\alpha$AMG. 
The setup update is done on the second time-scale for the shift $\mu+\rho_1$, since it is still close to the
light quark mass. In this case, we find slightly larger fluctuations in the iteration count of the outer level of the
multigrid solver compared to the $N_f=2$ simulation shown in Fig.~\ref{fig:iter_nf2}. Overall the simulation
is stable and we find no correlation of the iteration count with larger energy violations $\delta H$.

The solutions computed by DD-$\alpha$AMG for the shifted linear systems in the monomials $R_3$ and $R_4$ are accelerated by providing an initial guess. Considering a Taylor expansion, we obtain
\begin{equation}
(Q_{\ND} \pm i\delta)^{-1} \simeq Q_{\ND}^{-1} \mp i\delta Q_{\ND}^{-2}.
\end{equation}
Thus we can use the inversions of the previously computed shift, i.e.~$Q_{\ND}^{-1}$ and $Q_{\ND}^{-2}$, for constructing an appropriate initial guess for the next shift.
This saves up to 30\% of the inversion time.

The computational cost of each monomial per MD trajectory is depicted in Fig.~\ref{fig:hmc_nf2+1+1}. 
The standard approach depicted in the left panel involves the employment of multi-mass-shift CG (MMS-CG) for inverting all in 
once the shifted linear systems. In the right panel we have depicted the costs of the simulation accelerated by using the DD-$\alpha$AMG solver for the most ill-conditioned 
linear systems. We achieve a speed-up for the $N_f=1+1$ sector of a factor of 2 compare to a full eo-MMS-CG algorithm. The overall speed-up for $N_f=2+1+1$ simulation at the 
physical light, strange and charm quark mass is a factor of 5.
\begin{figure}
	\centering
	\includegraphics[width=0.8\linewidth]{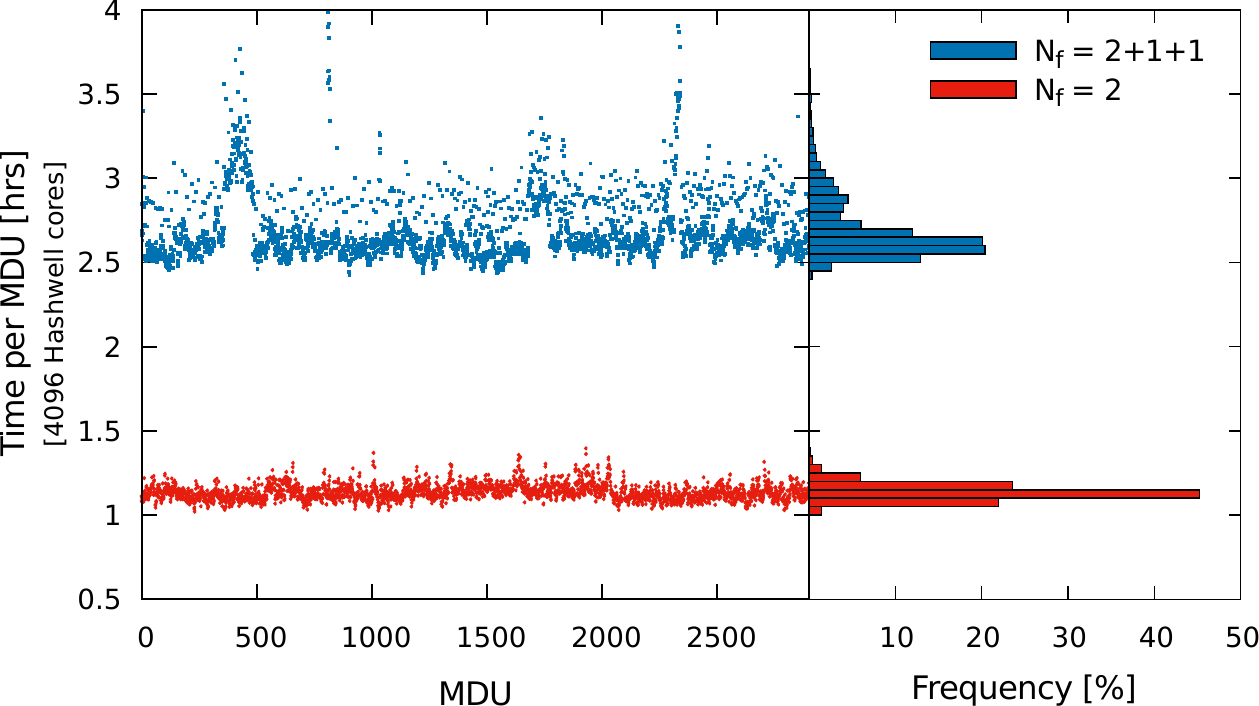}
	\caption{\label{fig:timenf211} Time per MDU of the $N_f=2$ and $N_f=2+1+1$ ensembles 
		in Table~\ref{tab:ensembles}. The $N_f=2+1+1$ timings also include the calculation of 
		the smallest and largest eigenvalue done at least every ten trajectories. The large fluctuation 
		around MDU 500 for the $N_f=2+1+1$ ensemble is due to a bad allocation in the machine 
		(it lasts exactly the duration of the allocation).}
\end{figure}
\section{Conclusions}

The multigrid accelerated simulations with TM fermions are discussed, showing a speed-up for simulations at the physical value of the pion mass of a factor of 8 in
the case of $N_f = 2$ and of a factor of $5$ in the case of  $N_f = 2 + 1 + 1$ compared to the  simulations performed without multigrid. 
We  use a three-level DD-$\alpha$AMG method optimized for TM fermions~\cite{Alexandrou:2016izb}. 
No instabilities in the iterations count are seen along the whole simulations.
As depicted in Fig.~\ref{fig:timenf211}, the time per MDU is  quite stable with 
fluctuations within the 10\%. During the $N_f=2+1+1$ simulation we calculate the smallest and largest eigenvalue of the non-degenerated TM operator at least every ten
MDUs, which takes additionally 15 mins. This explains the single points that are  frequently out of the main distribution. The two longer fluctuations instead are due to machine instabilities since they are limited to a single allocation of the job.
Although the multigrid method is limiting significantly the parallelization of the 
calculation, the speed-up  achieved makes feasible to sample enough
MD trajectory. 
Indeed, the average time per MDU in the $N_f=2$ simulation is slightly 
larger than an hour and in the $N_f=2+1+1$ simulation below three hours. 
In both cases, 4096 cores employing 147 Haswell-nodes on SuperMUC are used.
Indeed, we find that the three-level DD-$\alpha$AMG method shows an almost ideal scaling up to this number of cores
for a lattice with volume $V=64^3 \times 128$.

\subsection*{Acknowledgments}

This project has received funding from the  Horizon 2020 research and innovation program of the European Commission
under the Marie Sklodowska-Curie grant agreement No 642069. S.B.~is supported by this program.
This project has also received founding from PRACE Fourth Implementation Phase (PRACE-4IP) program of the European Commission
under grant agreement No 653838.
We would like to thank P. Dimopoulos, R. Frezzotti, B. Kostrzewa and C. Urbach for fruitful discussions during the generation of the ensembles.
The authors gratefully acknowledge 
the Gauss Centre for Supercomputing under project number~\emph{pr74yo}  for
providing computing time on the GCS Supercomputer SuperMUC at Leibniz Supercomputing Centre.

\bibliography{lattice2017}

\begin{thebibliography}{16}

\bibitem{Babich:2010qb}
R.~Babich, J.~Brannick, R.C. Brower, M.A. Clark, T.A. Manteuffel, S.F.
  McCormick, J.C. Osborn, C.~Rebbi, Phys. Rev. Lett. \textbf{105}, 201602
  (2010), \texttt{1005.3043}

\bibitem{Frommer:2013fsa}
A.~Frommer, K.~Kahl, S.~Krieg, B.~Leder, M.~Rottmann, SIAM J. Sci. Comput.
  \textbf{36}, A1581 (2014), \texttt{1303.1377}

\bibitem{Alexandrou:2016izb}
C.~Alexandrou, S.~Bacchio, J.~Finkenrath, A.~Frommer, K.~Kahl, M.~Rottmann,
  Phys. Rev. \textbf{D94}, 114509 (2016), \texttt{1610.02370}

\bibitem{Clark:2016rdz}
M.A. Clark, B.~Joó, A.~Strelchenko, M.~Cheng, A.~Gambhir, R.~Brower (2016),
  \texttt{1612.07873}

\bibitem{Frezzotti:2003ni}
R.~Frezzotti, G.C. Rossi, JHEP \textbf{08}, 007 (2004),
  \texttt{hep-lat/0306014}

\bibitem{Bacchio:2016bwn}
S.~Bacchio, C.~Alexandrou, J.~Finkenrath, A.~Frommer, K.~Kahl, M.~Rottmann, PoS
  \textbf{LATTICE2016}, 259 (2016), \texttt{1611.01034}

\bibitem{Finkenrath:2017}
J.~Finkenrath, C.~Alexandrou, S.~Bacchio, P.~Charalambous, P.~Dimopoulos,
  R.~Frezzotti, K.~Jansen, B.~Kostrzewa, C.~Urbach, \textbf{LATTICE2017} (2017)

\bibitem{Luscher:2003qa}
M.~L{\"u}scher, Comput. Phys. Commun. \textbf{156}, 209 (2004),
  \texttt{hep-lat/0310048}

\bibitem{Luscher:2007se}
M.~L{\"u}scher, JHEP \textbf{07}, 081 (2007), \texttt{0706.2298}

\bibitem{Jansen:2009xp}
K.~Jansen, C.~Urbach, Comput. Phys. Commun. \textbf{180}, 2717 (2009),
  \texttt{0905.3331}

\bibitem{Omelyan:2003272}
I.~Omelyan, I.~Mryglod, R.~Folk, Computer Physics Communications \textbf{151},
  272  (2003)

\bibitem{Abdel-Rehim:2015pwa}
A.~Abdel-Rehim et~al. (ETM), Phys. Rev. \textbf{D95}, 094515 (2017),
  \texttt{1507.05068}

\bibitem{Hasenbusch:2002ai}
M.~Hasenbusch, K.~Jansen, Nucl. Phys. \textbf{B659}, 299 (2003),
  \texttt{hep-lat/0211042}

\bibitem{Frommer:2013kla}
A.~Frommer, K.~Kahl, S.~Krieg, B.~Leder, M.~Rottmann (2013), \texttt{1307.6101}

\bibitem{Clark:2006fx}
M.A. Clark, A.D. Kennedy, Phys. Rev. Lett. \textbf{98}, 051601 (2007),
  \texttt{hep-lat/0608015}

\bibitem{zolotarev1877application}
E.~Zolotarev, Zap. Imp. Akad. Nauk. St. Petersburg \textbf{30}, 1 (1877)

\end{thebibliography}

\end{document}